\documentclass[prl,floatfix,twocolumn,showpacs,preprintnumbers,amsmath,amssymb,10pt]{revtex4}

\usepackage{graphicx}
\usepackage{dcolumn}
\usepackage{bm}

\begin{document}

\preprint{}

\title{Clustering and Micro-immiscibility in Alcohol-Water Mixtures:\\
Evidence from Molecular Dynamics Simulations}

\author{Susan K. Allison}
\author{Joseph P. Fox}%
\author{Rowan Hargreaves}%
\author{Simon P. Bates}%
\affiliation{%
School of Physics, University of Edinburgh,
Mayfield Road, Edinburgh EH9 3JZ 
}%

\date{\today}

\begin{abstract}
We have investigated the hydrogen-bonded structures in liquid methanol and a 7:3 mole fraction
aqueous solution using classical Molecular Dynamics simulations at 298K and ambient pressure. 
We find that, in contrast 
to recent predictions from X-ray emission studies, the hydrogen-bonded structure in 
liquid methanol is dominated
by chain and small ring structures. In the methanol-rich solution, we find evidence of 
micro-immiscibility, supporting recent conclusions derived from neutron diffraction data. 
\end{abstract}

\maketitle

\section{\label{sec:level1}Introduction}

The molecular structure and dynamics 
of hydogen-bonded networks is of fundamental 
importance across a wide range of scientific disciplines. 
The tetrahedral hydrogen-bonded 
structure of water is well-known and continues to be extensively
studied. 
Methanol, as the simplest alcohol, differs only in the 
presence of a methyl group in place of a proton, but is known to adopt a 
very different structure to water, comprising 
hydrogen-bonded chains and ring structures. 
The 
structure and
topology of the hydrogen-bonded
network in pure methanol and its aqueous solutions is an area where open questions still
remain as to the dominant structural motifs. Conclusions drawn from recent
experimental studies are conflicting \cite{guo0301,pali0001,dixit1}. This paper presents results of classical Molecular Dynamics (MD) simulations
that characterise the hydrogen-bonded clusters and the micro-immiscibility of the aqueous solution. 

Computer simulations have become increasingly used 
to aid predictions of structure and liquid methanol
is a good example of a widely-studied system. 
Investigations have ranged from simple pair 
potential calculations (e.g. \cite{guillot} and refs. therein),
through to the use of Monte Carlo methods to fit and interpret experimental 
data\cite{pali0001}, and now, 
to full {\em ab initio} studies of the structure and 
properties of the liquid \cite{pagl0301,meij0301}. In addition, hybrid schemes combining the 
rigour of first principles approaches and the expediency of pair potentials
have also been applied \cite{morr0301}.

Methanol is also the simplest amphiphilic molecule and its aqueous 
solutions are prototypical systems to study hydration of the hydrophobic
and hydrophilic moieties. 
It is structurally as simple as an
amphiphile can be, yet its aqueous solutions exhibit the characteristic
thermodynamic non-ideality of more complex systems. 
The traditional molecular-level 
description of hydration of hydrophobic species 
is enhancement of the 
structure of water immediately surrounding the hydrophobic group, leading to
ice-like or clathrate structures \cite{frank}. 
Within this model,
association of hydrophobic species is entropy-driven, liberating structured 
water. For decades, this has been
believed to explain the smaller than expected entropy change upon
mixing (based on what might be expected for ideally mixed solutions).
In recent years, this view has begun to be challenged by predictions from 
both computational \cite{laak9701} and experimental \cite{dixit1} studies investigating the structure and
properties of alcohol-water mixtures. 

Investigations of methanol-water mixtures have provided 
compelling counter-evidence to the traditional view of enhanced water 
structure around hydrophobic groups. 
MD simulations
by Meng and Kollman \cite{koll9601}, Laaksonen {\em et al} \cite{laak9701} 
and Fidler and Roger
\cite{fidl9901} point to preservation rather than enhancement of the water 
structure around the methyl group. 
Experimentally, Dixit {\em et al} \cite{dixit1} have 
employed the Empirical Potential Structure Refinement (EPSR) method to 
analyse data from neutron 
diffraction (ND) 
experiments and suggest the anomalous thermodynamics is due to 
incomplete mixing of water and methanol. This result is the most direct 
challenge to the `iceberg' model and suggests that rather than being enhanced
or depleted, the structure of water in the mixture is surprisingly close to 
that of the pure liquid. Further evidence of incomplete mixing has been 
presented very recently by Guo {\em et al} \cite{guo0301}, on the basis of X-ray 
emission (XE) spectroscopy. There are, however, apparent inconsistencies 
between these
experimental studies; the small fraction ($\sim$ 0.13) of the 
water molecules predicted to exist singly (i.e `singletons', with no
hydrogen-bonds to other water molecules) by Dixit {\em et al} is suggested to
be incompatible 
with the simulated XE spectra of Guo {\em et al}. 

The aims of this paper are twofold. Utilising classical MD
simulations, we investigate the cluster structures 
and their lifetimes in methanol, in order to provide insight into
conflicting predictions from recent experimental studies as to the dominant
structural motifs in the pure liquid. 
We also present results from simulations 
of a 
7:3 mole fraction 
aqueous solution, specifically aimed at 
providing molecular level insight into 
the predicted micro-immiscibility
of aqueous methanol.
On the basis of our analysis, we are able to 
predict the predominant hydrogen-bonded structures in pure methanol
and the changes these undergo upon hydration. In addition we 
show that a 7:3 aqueous solution does exhibit immiscibility on a 
microscopic scale, and we characterise the structure and lifetimes of 
water clusters within the solution. 

\section{\label{sec:details}Calculation details}

We have performed classical MD simulations 
within the {\em NVT} and {\em NPT} ensembles at 298K and ambient pressure. 
The classical simulations utilised 
previously tested potential parameters for both water \cite{levi9701} and
methanol \cite{pere0101,pere0201} that have been demonstrated 
to be give good predictions 
of the single component liquid structure and dynamics, at both ambient
and elevated pressures, in comparison to experimental data. 
Both molecules are modelled as fully
flexible entities, with van der Waals non-bonding interaction 
terms for each type of atom. Within the {\tt DLPOLY} 
\cite{dlpoly} program suite, we
have performed {\em NVT} 
simulations over 2ns duration with a timestep of 0.5fs on
 system sizes as follows: pure methanol, 512 molecules in a cubic box of size
33.03\AA: 7:3 mixture, 
297 methanol molecules and 127 waters, in a cubic box of edge 28.5\AA, corresponding to a density of 
the 7:3 mixture of 846 kgm$^{-3}$, in close agreement with the experimentally
measured density of 842 kgm$^{-3}$ \cite{dixit1}. 
Trajectory snapshots were saved
every 0.1ps for analysis. All simulations were equilibrated for 0.5ns prior
to data collection. 
Post-simulation analysis of the trajectories allowed calculation of the
self diffusion coefficients (via the Einstein relationship \cite{allentild})
and
site-site radial distribution functions (RDFs), $g(r)$.
%
%

The trajectory data was also used to characterise 
cluster distributions. The clustering algorithm was similar to that 
described previously to analyse data from the 
EPSR procedure \cite{dixit1}, originating from the formalism of Geiger 
{\it et al} \cite{geig7901}. 
A molecule is defined to be a member of a cluster if its
oxygen atom is found to be
within 
$R_{cut}$ of a neighbouring oxygen atom in the cluster 
(where $R_{cut}$ is the first minimum
in the oxygen-oxygen RDF, which is $\sim$3.5 \AA~ for both methanol and
water species). Defining a cluster in such a manner includes no angular 
prescriptions, and Pagliai {\it et al} \cite{pagl0301} have suggested 
tighter geometrical requirements, to eliminate unphysical 
clustering arrangements, including a HO--O angle smaller than 
30 degrees. Analysis of a representative fraction of clusters
predicted by our algorithm indicates that the overwhelming majority of 
clusters satisfy these additional constraints. 

An open question regarding the hydrogen-bonded structure in liquid methanol
is the suggested predominance of cyclic clusters \cite{guo0301}. We have calculated the 
proportion 
and distribution of sizes of cyclic 
clusters using an algorithm that iteratively `prunes' the
clusters previously identified to remove branches and non-cyclic 
structures. 
This procedure removes all linear and branched structures,
 leaving only cyclic structures which are either pure `n-mers' (single 
rings) or 
compound cycles comprising smaller cyclic structures. 

Cluster lifetimes were also deduced from the trajectory data, by calculating
the average duration that a molecule remains a participant in a 
cluster. A cluster is deemed to have `survived' between successive trajectory
snapshots if at least one member molecule remains the same. 
 The distinction is made between the 
lifetimes of singletons (a cluster size of 1) and
other cluster sizes. 

\section{\label{sec:results}Results}

Details of the structural predictions from the simulations of pure water and methanol 
have been reported elsewhere 
\cite{levi9701,pere0101}. Here we note only that 
in both cases, the bulk transport 
properties and the radial distribution functions (RDFs) are in 
good agreement with experimental data. 
However, in the case of pure methanol, we do present an analysis of 
the hydrogen-bonded network
 in the light of recent claims by 
Guo et al \cite{guo0301}, 
who claim 6-ring and 8-chain motifs to be the 
predominant structural feature in liquid methanol, in accordance 
with previous work \cite{sark9301} that had fitted the geometry of 
hexameric methanol units to X-ray diffraction data. Guo {\em et al} 
additionally claim
that ring and chain structures are present in equal abundance. 
This is in contrast to the predictions made by P\'{a}link\'{a}s {\em et al} 
\cite{pali0001} on the basis of reverse Monte Carlo fitting of 
ND data, employing a rigid, united atom structural model to describe 
methanol. This Monte Carlo study found no such predomination of hexameric 
rings; the average cycle size was predicted to be 4 molecules. 
In addition, it was found that cyclic structures comprise only
about one third of the total clusters of methanol, that 
the average length of hydrogen-bonded chains is also approximately 
4 molecules and 15\% of the methanol molecules exist as singletons. 

\begin{figure}
\resizebox{3.25in}{!}{\includegraphics{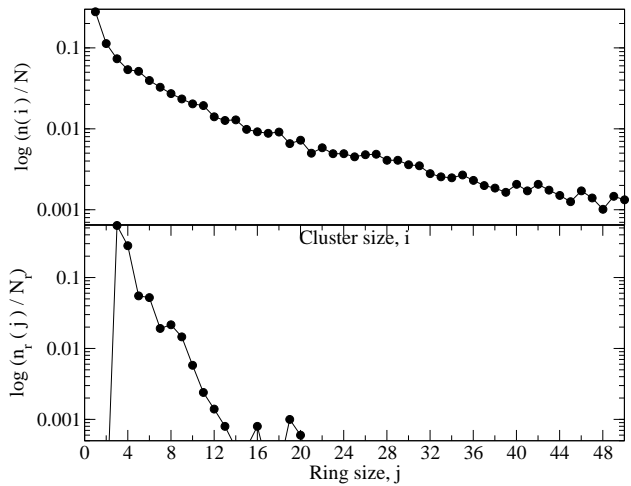}}
\caption{\label{fig:methchains} Clustering in pure methanol. The
figure shows the fraction of molecular clusters (upper panel) and rings
(lower panel).}
\end{figure}

The results of analysis of our MD simulations for clusters in pure
methanol is shown in Figure \ref{fig:methchains}. The 
number of clusters containing $i$ molecules is plotted as a 
fraction of total number of clusters, $m(i)/M$ (where $M=\sum_i m(i)$) against
the cluster size $i$. Of these clusters, we identify those containing 
cyclic structures, and plot the distribution of such clusters, $m_r(j)/M_r$,
as a function of ring size, $j$.  
We find that nearly 99\% of the methanol molecules exist in clusters of two 
molecules or more, but as a fraction of the total number of clusters, 
the singleton species account for a little over a quarter of all clusters found. The average cluster size 
is 23 molecules, but this is rather skewed by the presence of 
a small number of very large clusters. The figure shows that the 
proportion of clusters decreases rapidly with increasing cluster size. 
Analysis of 
cyclic structures present within clusters indicates that the most 
common cyclic structures are three and four membered rings, but pure rings of 
up to 20 molecules were found. 
These predictions are in good agreement with the reverse MC study
of P\'{a}link\'{a}s {\em et al} \cite{pali0001}. What discrepancies there 
are between their predicitions and our simulation results may be 
the result of a fixed geometry being used in the reverse MC study 
for representation of the methanol molecule. Visual inspection of 
some of our methanol clusters has shown that there is appreciable 
distortion of the local geometry around the methanol hydroxyl, particularly
if it is partcipating in more than one hydrogen bond, as found in the 
molecules in ring or branched chain structures. 
In contrast, our results do not support the recent claims of Guo {\it et al},
that 6- and 8-membered rings are the dominant structural 
features. We 
predict that $\sim$30\% of clusters 
contain one or more cyclic motifs (compared to the 50\% suggested by Guo {\it et al} 
and 33\% suggested by P\'{a}link\'{a}s {\it et al}). 
\begin{figure}
\resizebox{3.25in}{!}{\includegraphics{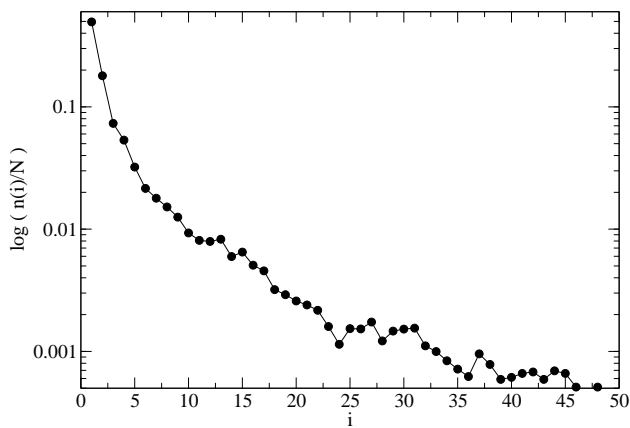}}
\caption{\label{fig:clustering} Clustering of water molecules in aqueous solution.
}
\end{figure}

Analysis of the lifetimes of the clusters indicate that on average 
15\% of the clusters are very short-lived, lasting not more than 1 or 2 ps,
whereas others lived for many hundreds of ps. 
Such an investigation of cluster size and lifetimes precludes the use of 
{\it ab initio} simulations to probe such information. The calculations
are simply too costly to be able to be performed on a statistically 
significant number of molecules for a long enough simulation time.

One might expect that a methanol-rich aqueous solution would be homogeneously
mixed, as methanol and water are known to be miscible across the entire 
range of liquid composition. 
Simulations on the 7:3 methanol-water aqueous solution reveal that the system
in fact comprises clusters of water, solvated by the hydrophilic ends of the 
methanol molecules forming the surrounding fluid.
The distribution of water cluster sizes in the mixture 
is shown in Figure \ref{fig:clustering}
 and the overall 
shape of the distribution appears to be relatively insensitive to system
size (results not reported here).
What we find is in close agreement with the results of Dixit {\em et al} 
on the basis of an EPSR fit to their neutron diffraction data.
The results show that water clusters exist within the system, sometimes
comprising up to 70\% of the water in the simulation box and that only
relatively few water molecules ($\sim$12\% over the duration of the 
simulation) are predicted to exist as single molecules, {\em i.e.} without
hydrogen bonds to other neighbouring water molecules.
This is the so-called
`free-swimming' water that is predicted by Guo {\em et al} to be incompatible
with their results. 

\begin{figure}
\resizebox{3.25in}{!}{\includegraphics{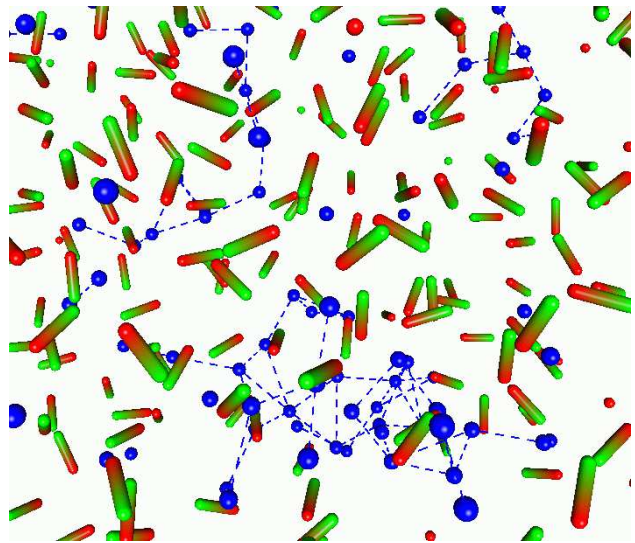}}
\caption{\label{fig:clusterpic} Visual snapshot of water clustering in 7:3 mixture. 
Spheres represent water oxygen atoms, sticks represent methanol carbon-oxygen bonds
and dashed lines indicate presence of a hydrogen bond between neighbouring water molecules. 
}
\end{figure}

A snapshot of a part of the simulation box, illustrating typical configurations is shown in 
Figure \ref{fig:clusterpic}; 
here only water
oxygens (spheres) and methanol carbon-oxygen bonds (sticks) are visible for simplicity.
The snapshot clearly shows the presence of a large globular cluster in the lower part of the 
image, and a more extended, bifurcated chain cluster in the upper part. 
The large globular cluster has
a radius of gyration 5.2 \AA~ and the average hydrogen
bond length in the cluster is $\sim$ 2 \AA. 

We have further analysed the simulation results to 
calculate an average lifetime of the singleton water
molecules. We find that although they account for 12\% of all water
molecules over the duration of the simulation, they are very short-lived. On
average, each single water molecule exists in an unbound state for only a few 
ps of the simulation. Occasionally, one persists for as long as a
few tens of ps, but this is rare. This lifetime is consistent 
with a picture of dynamic water clusters, frequently shedding and re-absorbing
individual water molecules. In contrast, the lifetimes of the larger clusters
can be much longer. These can persist in the simulation for hundreds of ps,
principally because of the larger cluster sizes. We note that the calculated diffusion 
coefficients of the water and methanol species in the mixture (1.5 $\times 10^{-9} m^2s^{-1}$ and
1.9 $\times 10^{-9} m^2s^{-1}$, repsectively) are both smaller than the corresponding values for 
the pure components (2.4 $\times 10^{-9} m^2s^{-1}$  and 2.6 $\times 10^{-9} m^2s^{-1}$).

The presence of the water clusters disrupts the hydrogen bond network of methanol molecules. 
In the mixtures, we find almost double the fraction of methanol clusters existing as 1, 2, or 3
member clusters. If we define a `maximum' cluster size as one that accounts for at least 1\% of
all clusters of a given species, 
we find that this threshold is reached at a methanol cluster size of 10 in the mixture 
simulations, compared with a value of 17 in the pure liquid simulations. The distribution of
ring sizes in the methanol clusters in the aqueous solution is similarly affected; 80\% of all structures
comprise three membered rings and there are no rings larger than hexamers. There is a marked
decrease in the percentage of methanol clusters that include a ring structure; this was found to be 
$\sim$ 40\% in the pure liquid simulation, but drops to $\sim$ 10\% in the mixture. 
The average lifetime of such clusters is also much shorter compared to those in the 
pure liquid; in the mixture, the majority of methanol clusters persist for 1ps or less 
and no cluster survives longer than 40ps.

\begin{figure}
\resizebox{3.25in}{!}{\includegraphics{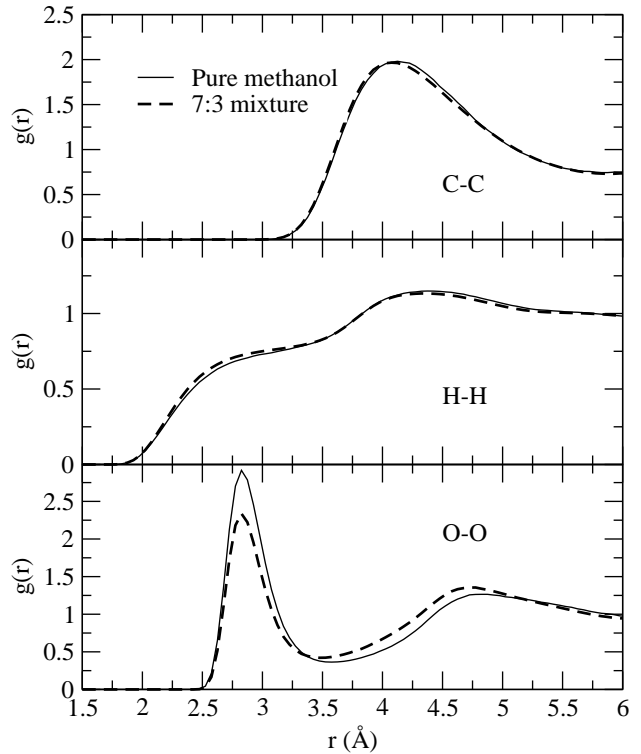}}
\caption{\label{fig:rdfs} Site-site radial distribution functions for carbon,
methyl hydrogen and oxygen atoms of methanol in the aqueous solution and the pure liquid.
}
\end{figure}

The intercalation of water into the methanol network is not the only reason for this dramatic
change in the hydrogen bonded structures; a compounding reason is that the methyl headgroups
of neighbouring methanol molecules are pushed closer together. 
The evidence for this methanol-methanol  association may be probed via the site-site
radial distribution functions ($g_{\alpha\beta}$) for methanol in the 7:3 mixture, in 
comparison with the same functions in the pure liquid. The $g_{CC}$, $g_{HH}$ (methyl
hydrogen) and $g_{OO}$ functions are shown in Figure \ref{fig:rdfs}. The RDFs for 
methanol in the mixture exhibit similar features to those described 
by Dixit {\em et al}, namely
the first peak in the C-C and the H-H RDFs
shift inward slightly. The latter is not found to move as much as seen from the 
EPSR fitting to experimental results, and this is possibly due to differences in 
the way internal degrees of freedom are treated in the MD simulations and 
the EPSR procedure. The first peak of the O-O function diminishes 
in size and the second peak moves to lower $r$. These RDFs are consistent with 
the picture that the methyl groups are compressed in the 7:3 mixture and the hydroxyl
groups pushed slightly further apart. 
Analysis of the corresponding
 RDFs for the water species in the mixture reveal that the water structure 
bears many of the signatures of pure water; the peaks in the site-site radial distribution
functions are similar to those from the simulations of the pure liquid. 

In conclusion, we have applied classical MD simulations to study the structure and dynamical 
properties of hydrogen-bonded clusters in pure methanol and a 7:3 mole fraction aqueous solution. 
Contrary to recent claims, we find no evidence of large cyclic structures forming the predominant
structural motif in liquid methanol. We predict that the aqueous solution exhibits 
immiscibility on a microscopic scale, with methanol molecules associating via closer
methyl group contact. Work is currently in progress to investigate the temperature and 
pressure dependence of these phenomena. 
\begin{acknowledgments}
We gratefully acknowledge the support of IBM and EPCC for the provision of computing 
facilities used for this work and the EPSRC for grant funding to SA, JF and RH.
\end{acknowledgments}

\end{document}